\documentstyle[aps,preprint,prc,12pt]{revtex}
\begin{document}
\baselineskip 24pt
\vspace{0.5 truecm}

\begin{center}
{\Large \bf  Isotopic Composition of Fragments in Nuclear 
Multifragmentation}
\end{center}
\vspace{0.4 truecm}
\begin{center}
{\large P.M. Milazzo$^{1}$, A.S. Botvina$^{2,3}$, G. Vannini$^{2}$, 
N. Colonna$^{4}$, F. Gramegna$^{5}$, G. V. Margagliotti$^{1}$, 
P.F. Mastinu$^{5}$, A. Moroni$^{6}$, R. Rui$^{1}$ }
\end{center}
\vspace{0.4 truecm}

{\it 
$^{1}$Dipartimento di Fisica and INFN, 34127 Trieste, Italy

$^{2}$Dipartimento di Fisica and INFN, 40126 Bologna, Italy

$^{3}$Institute for Nuclear Research, Russian Academy of Science, 
 117312 Moscow, Russia

$^{4}$INFN, 70126 Bari, Italy

$^{5}$INFN, 35020 Laboratori Nazionali di Legnaro, Italy

$^{6}$Dipartimento di Fisica and INFN, 20133 Milano, Italy}

\vspace{0.3 truecm}
\begin{abstract}
The isotope yields of fragments, produced in the decay of the 
quasiprojectile in Au+Au peripheral collisions at 35 MeV/nucleon 
and those coming from the disassembly of the unique source formed in 
Xe+Cu central reactions at 30 MeV/nucleon, were measured.
We show that the relative yields of neutron-rich isotopes increase with the 
excitation energy in multifragmentation reaction. In the framework of 
the statistical multifragmentation model which fairly well reproduces the 
experimental observables, this behaviour can be explained by 
increasing $N/Z$ ratio of hot primary fragments, that corresponds to 
the statistical evolution of the decay mechanism with the excitation energy: 
from a compound-like decay to complete multifragmentation. 
\end{abstract}
\vspace{0.2 truecm}

\hspace{0.5 truecm}PACS numbers: 25.70Pq, 25.70-z, 24.60-k
\newpage


Nuclear fragmentation and its connection to the behaviour of nuclear 
matter at high excitation energy is the subject of intensive theoretical and 
experimental investigations \cite{hirschegg}. Some general properties of 
this process are already established: at relatively 
small excitation energies ($E^*\leq$2--3 A~MeV) there is a formation and decay 
of a long--lived  compound-like nucleus system. This process can be described 
by evaporation/fission--like models. At higher excitation energies (close to 
the binding energy) there is a complete fast disintegration of the system into 
fragments. In this case statistical models based on the hypothesis 
of a nuclear phase transition (simultaneous decay) happen to be 
very successful \cite{PR,gross}.

The 
parameters of the nuclear system at the break-up have been studied by many 
methods (see e.g. \cite{deses}). 
Information about the density in the freeze-out volume is usually extracted 
from the analysis of velocity correlation functions and kinetic energies of 
fragments. The temperature of the system can be deduced from different 
observables: 1) relative populations of unstable nuclear levels, 
2) fragments' kinetic energies, or 3) relative production of isotopes. 
The last method, based on the statistical properties of double isotope ratios 
\cite{albergo}, seems to be the most reliable one, and in fact for the first 
time it made possible to obtain a nuclear caloric curve as an experimental 
evidence of a nuclear liquid-gas type phase transition \cite{pochod}. 

Usually these methods need a correction for secondary decay of hot fragments 
produced in the freeze-out volume \cite{nayak}. It has to be stressed that 
the information about chemical composition of hot fragments is of primary 
importance: depending on their $N/Z$ ratio, the decay can in fact proceed 
differently. Moreover the knowledge of the chemical composition of the primary 
fragments would allow to 
more precisely establish the thermodynamical conditions at the freeze-out, 
e.g., to provide a way to apply the energy balance \cite{dagostino99}, 
as well as it allows to obtain 
unambiguous data to extract excitation energies of fragments via 
correlation analysis \cite{nmarie}. Furthermore it can hint 
for behaviour of the symmetry energy term of the nuclear Equation of State at 
subnuclear densities and make difference between dynamical and statistical 
mechanisms of fragmentation \cite{larionov99}.

In this paper we present recent data on the isotope production in heavy-ion 
collisions at intermediate energies \cite{milazzo98,milazzo99} with the aim of 
studying the isotope content of fragments for different sources 
and excitation energies during the transition from the low energy decay 
to the multifragmentation. We'll show that 
the behaviour of the experimental isotope yields as 
a function of source size, isospin and excitation energy can be 
connected to the corresponding evolution of the $N/Z$ ratio of 
the hot fragments, leading to more insight in the freeze-out 
condition. 


We investigated the Xe+Cu at 
30 MeV/nucleon and Au+Au at 35 MeV/nucleon reactions. The experiments
were performed at the National Superconducting K1200 Cyclotron
Laboratory of the Michigan State University. 
The angular range 3$^{\circ}<\theta_{lab}<$23$^{\circ}$ was covered by the
$MULTICS$ array \cite{mcs}.
The identification thresholds in the $MULTICS$ array were about 1.5
MeV/nucleon for charge identification and about 10 MeV/nucleon for 
mass identification. 
The MULTICS array consisted of 48 telescopes, each of which 
was composed of an
Ionization Chamber (IC), a Silicon position-sensitive detector (Si) 
and a CsI crystal. Typical energy resolutions 
were 2\%, 1\% and 5\% for IC, Si and CsI, respectively. 
Light charged particles
and fragments with charge up to Z=20 were detected at
23$^{\circ}<\theta_{lab}<$160$^{\circ}$ by the phoswich detectors of
the MSU $Miniball$ hodoscope \cite{mb}. The charge identification
thresholds were about 2, 
3, 4 MeV/nucleon in the $Miniball$ for Z= 3, 10, 18, respectively. 
The geometric acceptance of the combined array was
greater than 87\% of 4$\pi$. 


The multiplicity of detected charged particles (Nc) was used for the 
reduced impact parameter $\hat b$ reconstruction: 
$$\hat b = b/b_{max} = 
\left(\int_{Nc}^{+\infty} P(N'c) dN'c \right)^{1/2}.$$ 
Here P(Nc) is the charged particle 
probability distribution and $\pi\cdot b_{max}^2$ is 
the measured reaction cross section for Nc$\ge$3. 

The decay products coming from the decay of the quasiprojectile in peripheral
Au+Au 35 MeV/nucleon reaction and those coming from the disassembly of 
the unique source formed in Xe+Cu 30 MeV/nucleon central collisions 
have been identified through a careful data selection taking into 
account for the experimental efficiency distortions on energy and angular 
distributions \cite{milazzo98,milazzo99}. 
In particular it has been verified that all the detected decay products
are emitted nearly isotropically from the same source and that their 
energy distributions have Maxwellian shapes, i.e. that angular and 
energy distributions are compatible with a statistical emission, 
providing an experimental indication that thermalization has been 
reached \cite{milazzo98,milazzo99}. 
The reconstruction of excitation energies of the sources was carried out 
analyzing kinematic characteristics of the produced fragments 
(calorimetric evaluation \cite{dagostino99,calo}). 

We studied fragments coming from the excited quasiprojectile Au-like sources 
produced in 
peripheral Au+Au collisions with impact parameters 0.95$> \hat b>$0.5, 
corresponding approximately 
to excitation energies from 3 to 6 MeV/nucleon. Also we considered 
central ($\hat b<$0.2) Xe+Cu collisions which provide sources with 
excitation around 5.5 MeV/nucleon with approximately the same number of 
nucleons as Au but with larger charge (the thermal source can be considered as 
a system after total fusion of colliding nuclei in the center of mass system). 
In ref. \cite{milazzo98,milazzo99} these data were used to obtain a caloric 
curve. The double isotope ratio method allows to eliminate the need of the 
knowledge of the initial $N/Z$ value of the source, and provides estimates 
for the temperature. To get information about the composition of hot 
fragments, new observables should be studied, such as 
the isotope yields for fixed elements and their evolution with the excitation 
energy and other parameters of the emitting source. We think that 
an analysis of the relative isotope production can provide a more 
reliable information about statistical picture of the process than an 
analysis of the isobars. The neighbouring isobars (with $\Delta$Z=1) might be 
produced at different Coulomb barriers, the difference being up to 10 MeV for 
the Au source. The uncertainty in accounting the real Coulomb energy 
of the isobars in the freeze-out may essentially exceed the difference in 
their binding energy ($\sim$1 MeV) that prevents to conclude 
unambiguously about thermodynamical parameters. 

To avoid a possible problem of preequilibrium in the emission of light charged 
particles we concentrate mainly on IMFs in this study. In fig.~1 we show 
the relative isotope yields versus excitation energy obtained from the 
experimental data for Au sources (each isotope yield is normalized to the 
total yield for fixed $Z$ value). The relative yields change a little 
as function of the energy, however, one can see a very interesting 
feature: the relative yields of isotopes with big 
$N/Z$ ratios become larger with increasing excitation energy. 
In fig.~2 we present the ratios of yields of measured isotopes 
with the biggest and smallest number of 
neutrons at fixed $Z$ values versus the excitation energy. 
For all analysed IMFs, the ratio increases considerably in the energy range of 
$E^*$=3--6 MeV/nucleon. Within such range, the lowest energy corresponds to the 
onset of multifragmentation with mean IMF multiplicity around one plus 
a heavy residual, while at the highest energy the fragmentation 
into many IMFs dominates. Looking at the sources with different $N/Z$ 
ratios we found that the abundances of neutron-rich 
isotopes are larger for the peripheral quasiprojectile Au than for the 
central Xe+Cu unique source, 
at the same excitation energy of about 5.5 MeV/nucleon. 
The relative yields at this excitation energy are shown in fig.~3. 
This trend has a natural explanation as the $N/Z$ ratio 
of the Au source is larger than the Xe+Cu one. 

If we assume that 
the yields of the observed isotopes are mainly affected by the secondary decay 
of hot primary fragments having a slightly larger size, one can consider the 
presented data for the Au source as an experimental evidence 
that the $N/Z$ ratio of intermediate mass primary fragments increases with 
the excitation energy of the source. 
However the physical process behind of this evolution needs a clarification. 


It was shown that the statistical multifragmentation model (SMM) 
\cite{PR} very well reproduces the observed charge yields and 
the $He$-$Li$ isotope temperatures \cite{milazzo98,milazzo99}, as well as the 
mean fragment kinetic energies. 
In the following we use the set of SMM parameters which gives the 
best description of multifragmentation of the sources produced in peripheral 
collisions \cite{milazzo98,dagostino99}. The freeze-out 
density was taken 1/3$\cdot \rho_0$ ($\rho_0$ is the normal nuclear density). 
The crucial condition for the present isotope analysis is the requirement 
of a full description of the charge yields for each considered charge 
distribution. Under this condition the same parameterization for 
the central Xe+Cu collisions was used. A possible slight decrease of 
the source size, as a result of preequilibrium emission, does not affect the 
conclusions because it hardly changes the $N/Z$ ratio of the 
source \cite{PR}; likewise a small dynamical expansion effect has minor 
importance. 
We have checked that the calculated isotope trends 
(see below) remain stable with respect to reasonable variations of the SMM 
parameters in the ranges where the charge yields and other observables are 
reproduced. 

We performed a detailed analysis of the isotope production in the framework of 
this statistical model. Comparison 
of the SMM predictions with the data is shown in fig.~1, 2 and 3. 
The qualitative agreement is evident (even quantitative for some important 
isotopes) and the general trends, especially, the increase with excitation 
energy of the neutron-rich isotopes with respect to the neutron-deficit ones, 
are correctly reproduced. 
However there are few discrepancies in the results which require some 
justification. 
In the SMM the Fermi-break-up model is used to describe the secondary decay 
of fragments with $A\leq$16 \cite{botvina1}. It takes into account all ground 
and nucleon-stable excited states of light fragments and calculates the 
probabilities of population of these levels microcanonically (according to the 
available phase space). It does not include matrix elements of the 
transitions between these states that can be important at small excitation 
energies. Also the model does not take into account for possible shifts of the 
nuclear states caused by Coulomb interaction of the excited fragments with the 
surrounding nuclear matter: these shifts should be calculated in consistent 
quantum theories. 
However, in our cases we have a rather high excitation energy of hot 
primary IMFs: from 2 to 3 A~MeV and higher, that is considerably larger than 
thresholds of the main break-up channels, and the above mentioned problems 
do not affect the calculated trends. 
Obviously one can better see the $N/Z$ effect 
in yields of nuclides far from $\beta$-stability line, which are less 
influenced by the shell structure.  

According to the SMM predictions \cite{PR} at the beginning of 
multifragmentation the number of primary nucleons in the freeze-out is very 
small and nearly all available 
protons and neutrons are bound in hot primary fragments. Their $N/Z$ ratio 
depends on the fragment size. If there are light and heavy fragments in 
the freeze-out, the light fragments have typically smaller 
$N/Z$ ratio than the heavy ones: these channels are more energetically 
favorable because of an interplay of the symmetry and Coulomb energies. 
In fig.~4 we show how the mean $N/Z$ 
ratios for the light ($Z=8-10$) and heavy ($Z=68-70$) primary fragments 
evolve with the excitation energy. It is interesting to note that, if a big 
(residue-like) hot fragment is present in the freeze-out, its $N/Z$ ratio 
can be even larger than the corresponding source ratio because the other light 
fragments have a considerably lower ratio. At excitations around 
the multifragmentation threshold ($E^*_{thr}=$3-4 MeV/nucleon) big fragments 
start to disappear and nearly all available neutrons are combined into hot IMFs 
giving rise to their neutron content. 
Finally at very high excitations ($E^*\sim$8 MeV/nucleon) the 
$N/Z$ ratio of the hot IMFs starts to decrease because the number of 
primary free neutrons increases fastly. 
This behaviour is responsible for the corresponding trends of the cold 
fragments produced after deexcitation of these primary IMF (see in 
Fig.4, e.g., Z=5). 
Mainly as a consequence of this evolution, we observe an increase of the $N/Z$ 
ratio of the cold fragments in the energy range $E^*$=3--6 
MeV/nucleon, and a sizeable change in the relative yields of neutron-rich 
and neutron-deficit isotopes (see figs.~1, 2). At higher energies this ratio 
should drop similarly to the hot fragment one. 

We should point out that different mechanisms are responsible for the fragment 
production in the SMM. At small excitation energy light IMFs can be 
emitted from a compound-like nucleus system. 
It favours the production of nearly symmetric isotopes with 
large binding energy (close to $\beta$-stability line). Therefore their 
$N/Z$ ratios are usually smaller than the ratio of the source. The probability 
of the evaporation of IMF is small, however, it contributes to the yield at 
$E^* \leq E^*_{thr}$. At excitation energies higher than the threshold the 
multifragmentation sets in: from a fast break-up into 
two hot fragments it evolves towards the break-up into three or more fragments 
with the increase of the source excitation energy \cite{PR}. 
At the multifragmentation the secondary decay of hot primary 
fragments is the main process defining a relative abundance of particular 
isotopes. In the SMM calculations the contribution of the evaporated 
isotopes favours the low ratios presented in fig.~2 at 
small $E^*$. However, assuming that only the evaporation mechanism 
exists at high 
excitations (independent from the reproduction of the charge yield) 
the observed increase of the ratio can not be explained. That supports the 
suggested evolution of the decay mechanisms and isotope composition of hot 
fragments.

It was shown in Ref.\cite{xutsang} that an increase of the $N/Z$ source 
ratio leads to increasing the relative yields of neutron-rich isotopes, 
in agreement with present results. 
In their analysis they extract also information about 
neutron-to-proton ratio ($n/p$) at the freeze-out. In the present calculations 
with SMM we found that for central (Xe+Cu) and peripheral (Au) sources 
with the excitation of 5.5 A~MeV this ratio increases 
more than increasing the $N/Z$ ratio of the corresponding sources, 
in agreement with \cite{xutsang}. 
A possible production of neutron-rich hot primary fragments is also predicted 
by other theoretical models. Dynamical stochastic mean field calculations 
\cite{larionov99} for Au source predict hot fragments with 
the same $N/Z$ ratio as the SMM. Also an analysis of the correlation 
functions performed in \cite{nmarie} comes to the conclusion that the larger 
neutron content of the hot primary fragments corresponds to the experimental 
data. 

In summary, we presented new data on yields of isotopes produced after decay 
of the Au and Xe+Cu sources at excitation energy range of 3--6 MeV per 
nucleon, that is around and slightly 
above the multifragmentation threshold. 
We found that the experimental relative yields of neutron--rich 
isotopes increase with excitation energy for the Au sources. 
The SMM calculations reproduce the whole set of data well enough to 
support the statistical picture realized in this model. 
In this approach the energy dependence of the isotopic composition of the 
produced fragments can be explained in terms of a transition from 
an evaporation-like emission of few fragments to the total 
multifragmentation break-up which leads to the increase of neutron content 
of hot primary intermediate mass fragments.

\vspace{0.5 truecm}
A.S. Botvina thanks the 
Istituto Nazionale di Fisica Nucleare (Bologna section, Italy) for hospitality 
and support.

\vspace{0.3 truecm}
{\large \bf  {Figure captions}}\\

{\bf Fig.1:} {Relative yields of isotopes of different elements 
versus excitation energy of Au source. Symbols are experimental data, 
lines are SMM calculations: 
$^6$Li, $^7$Be, $^{10}$B, $^{11}$C (open circles, dashed line); 
$^7$Li, $^9$Be, $^{11}$B, $^{12}$C (full circles, solid line); 
$^8$Li, $^{10}$Be, $^{12}$B, $^{13}$C (open squares, dot-dashed line); 
$^{14}$C (full squares, dotted line). 
The experimental uncertainties on the excitation 
energy are the same as shown in fig.~2; error bars on relative yields are 
smaller than symbols size. 
}
\vspace{0.5 cm}

{\bf Fig.2:} {Ratio of relative yields of neutron-rich to neutron-deficit 
isotopes of Li, Be, B and C fragments versus excitation energy of Au source. 
Solid circles are the experimental data, while lines refer to SMM calculations;
solid squares refer to central Xe+Cu experimental data. }
\vspace{0.5 cm}

{\bf Fig.3:} {Relative yields of different isotopes for fragments with charges 
from $Z$=1 to $Z$=6. Circles are experimental data: the solid ones are for the 
Au system, the open ones are for the Xe+Cu system at the excitation energy 
of 5.5 MeV/nucleon. Solid and dashed lines are the corresponding SMM 
calculations.}
\vspace{0.5 cm}

{\bf Fig.4:} {The SMM calculations of the mean neutron--to--proton ($N/Z$) 
ratio 
of hot primary fragments produced at the freeze-out (full lines) 
and the cold fragments 
produced after the secondary decay (dot-dashed line) versus excitation energy 
for the Au source.}


\begin{thebibliography}{99}
\baselineskip=0.3 pt

\bibitem{hirschegg} See for instance: 
"Multifragmentation", Proceedings of the International 
Workshop 27 on Gross Properties of Nuclei and Nuclear Excitations, 
Hirschegg, Austria, January 17--23, 1999. GSI, Darmstadt, 1999. 
\bibitem{PR} J.P. Bondorf, A.S. Botvina, A.S. Iljinov, I.N. Mishustin and 
K. Sneppen, Phys. Rep. {\bf 257},133 (1995).
\bibitem{gross} D.H.E. Gross, Rep. Progr. Phys. {\bf 53}, 605 (1990).
\bibitem{deses} P. Desesquelles et al., Nucl.Phys. {\bf A633}, 547 (1998).
\bibitem{albergo} S. Albergo et al., Nuovo Cimento {\bf 89},1 (1985).
\bibitem{pochod} J. Pochodzalla et al., Phys. Rev. Lett. {\bf 75}, 1040 (1995).
\bibitem{nayak} T.K. Nayak et al., Phys. Rev. {\bf C45}, 132 (1992).
\bibitem{dagostino99} M. D'Agostino et al., Nucl.Phys. {\bf A650}, 329 (1999).
\bibitem{nmarie} N. Marie et al., Phys. Rev. {\bf C58}, 256 (1998).
\bibitem{larionov99} A.B.Larionov, A.S.Botvina, M.Colonna and M.Di Toro. 
Nucl. Phys. {\bf A658}, 375 (1999). 
\bibitem{milazzo98} P.M. Milazzo et al., Phys. Rev. {\bf C58}, 953 (1998).
\bibitem{milazzo99} P.M. Milazzo et al., Phys. Rev. {\bf C60}, 044606 (1999).
\bibitem{mcs} I. Iori {\it et al.}, Nucl. Instr. and Meth. Phys.
Res. {\bf A325}, 458 (1993). 
\bibitem{mb} R. T. de Souza {\it et al.}, Nucl. Instr. and Meth.
Phys. Res. {\bf A295}, 109 (1990). 
\bibitem{calo} X.Campi et al., Phys. Rev. C {\bf 50}, R2680 (1994).
\bibitem{botvina1} A.S. Botvina et al., Nucl. Phys. {\bf A475}, 663 (1987).
\bibitem{xutsang} H. S. Xu {\it et al.}, NSCL-MSU Report 1137, oct. 1999.

\end{thebibliography}
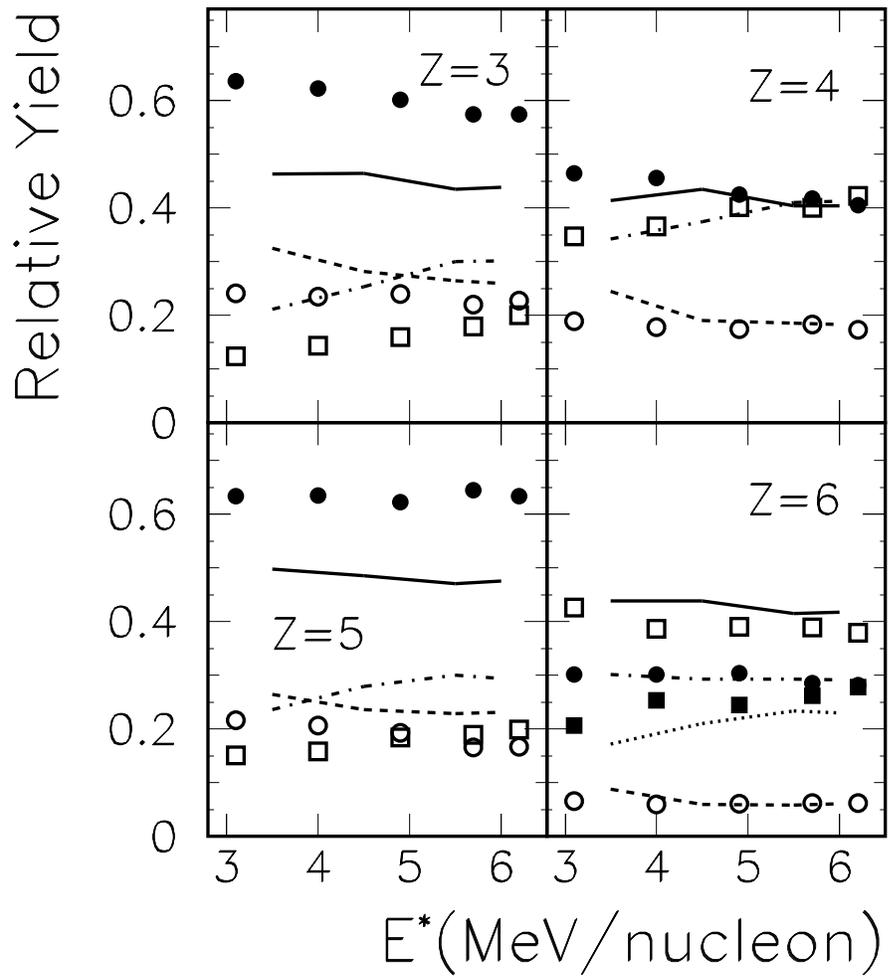
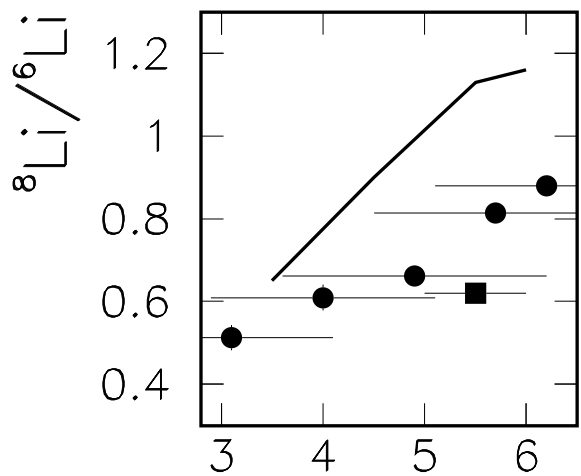
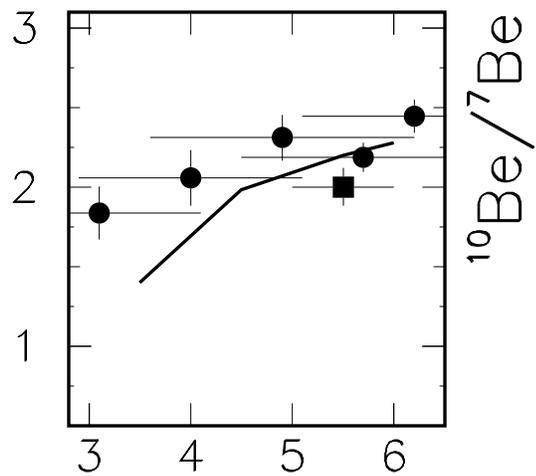
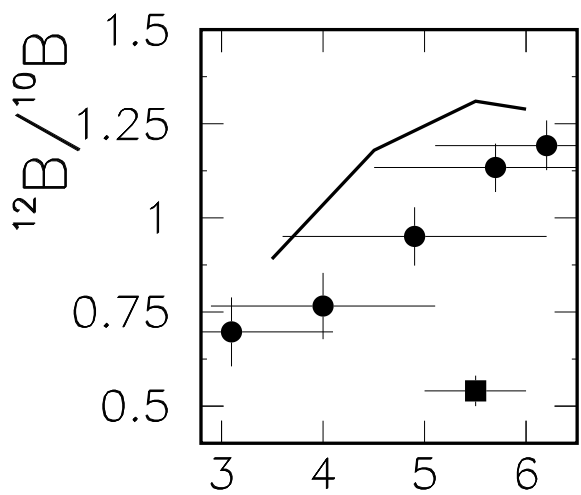
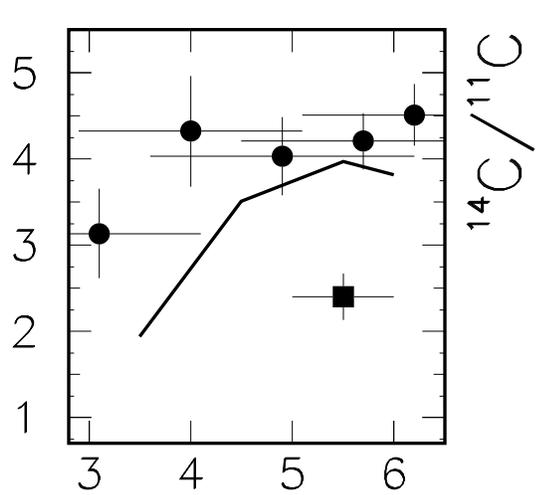
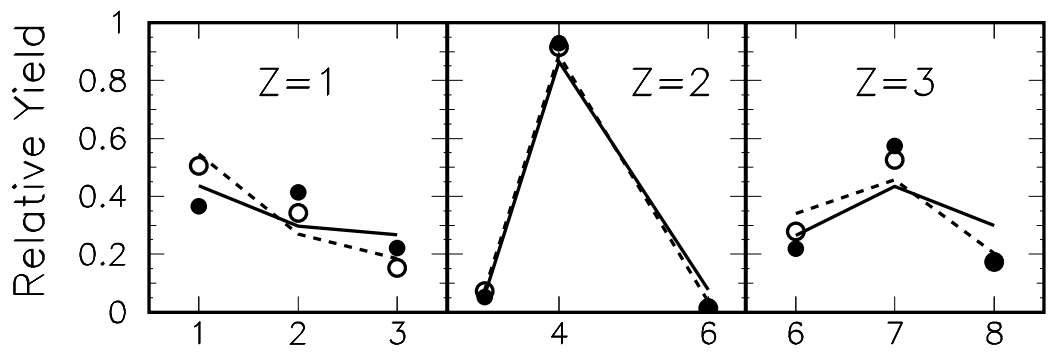
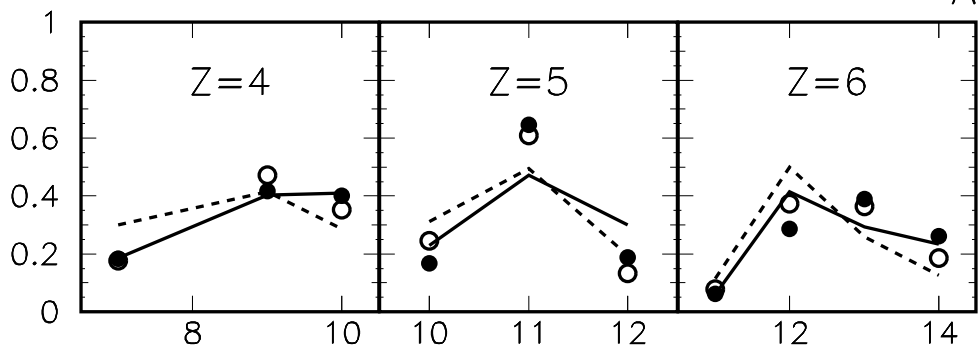
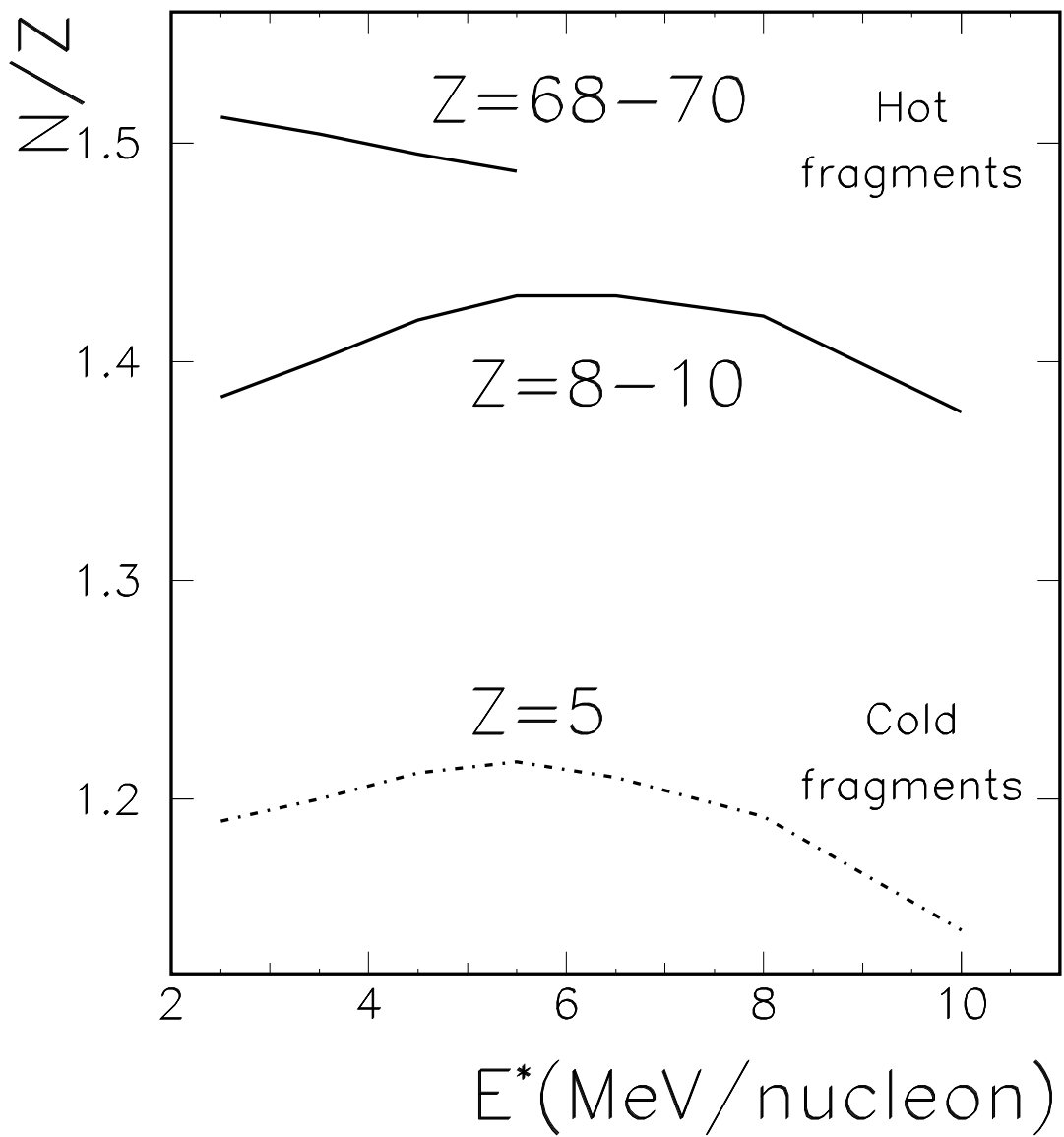
\end{document}